\documentclass[12pt]{article}
\usepackage{amsmath,amssymb,graphicx,mathrsfs,hyperref,slashed}

\usepackage{graphicx,picture,cleveref,float,caption}
\usepackage{subcaption}
\usepackage{geometry}

\renewcommand{\baselinestretch}{1.5}
\newcommand{\avg}[1]{\langle #1 \rangle} 

\crefname{equation}{Eq.}{Eqs.}
\crefname{figure}{Fig.}{Figs.}
\crefname{section}{Section}{Sections}
\usepackage{tensor}

\author{\large M.M.W. Shawa$^{(1)}$,  A.J.M. Medved$^{(1,2)}$
\\
\vspace{-.5in} \hspace{-1.5in} \vbox{
 \begin{flushleft}
$^{\textrm{\normalsize (1)\ Department of Physics \& Electronics, Rhodes University,
  Grahamstown 6140, South Africa}}$
$^{\textrm{\normalsize (2)\ National Institute for Theoretical Physics (NITheP), Western Cape 7602,
South Africa}}$
\\ \small \hspace{1.07in}
  markshawa@aims.ac.za,\  j.medved@ru.ac.za
\end{flushleft}
}}
\date{}
\title{$N$-point functions of strongly coupled fluids dual to generalized higher-derivative theories of gravity}
\begin{document}

\maketitle
\abstract{We calculate  the connected stress-tensor correlation functions that are dual to a certain class of  graviton scattering amplitudes in an asymptotically anti-de Sitter, black brane spacetime. This is a continuation of a previous study in which one-particle-irreducible amplitudes were calculated for arbitrarily higher-derivative gravity in a particular kinetic regime of high energies and large scattering angles. The utility of this regime is twofold: It is particularly well suited  for translating  scattering amplitudes  into the language of the gauge theory and 
it emphasizes the contributions from higher-derivative corrections (which would
otherwise be perturbatively suppressed). Using the new results, we  show how it could be possible to identify, experimentally, the gravitational dual to
a  strongly coupled fluid such as the  quark-gluon plasma.}
\renewcommand{\baselinestretch}{1.5}\normalsize

\section{Introduction}
The gauge--gravity duality provides a framework for learning about
strongly coupled,  $D$-dimensional gauge theories by studying theories of weakly coupled, $(D+1)$-dimensional gravity~\cite{Maldacena}. One of the duality's main ``selling points''  is that  strongly coupled field theories ---  which are, in general, 
 notoriously difficult to analyze --- can now  be readily investigated. Here,
we will be particularly interested in the  correspondence between  
stress-tensor correlation functions on the gauge side and  graviton scattering amplitudes on the side of the gravitational dual~\cite{Hofman,Witten}.

Back in the early days of this program, much attention was given to five-dimensional anti-de Sitter (AdS) space and four-dimensional super Yang--Mills  theory. Taking the limit $\lambda, N \to \infty$  while $g_s \to 0$, one arrives at a theory that corresponds to  pure Einstein's gravity~\cite{Aharony}. (Here, $\lambda = g^2_sN$ is the `t Hooft coupling of the gauge theory, $N$ is its rank and $g_s$ is the coupling strength of the implicit string theory.) An immediate consequence of this limit is that, in a $\lambda^{-1}$ expansion, loop diagrams are suppressed and a tree-level analysis is sufficient. But suppose that one is dealing with a physical system which is described by a gauge theory of 
large-yet-finite rank. Then the gravitational dual would, as one might anticipate, include  a multi-derivative correction to Einstein's (two-derivative) theory ({\em e.g.}, \cite{Buchel:2008vz}). 

In this paper, we start with the graviton scattering amplitudes for generic, multi-derivative theories of gravity (with or without matter fields). These scattering amplitudes must be gauge-invariant quantities and can, by extension, be constructed from a gauge-invariant action. In general, a multi-derivative theory of gravity can be described by a Lagrangian that consists of a series of invariant products of the four-index Riemann tensor and its contracted forms,
the  Ricci tensor and scalar.~\footnote{We are, for simplicity, overlooking the case in which covariant derivatives appear explicitly in the action.} However, as demonstrated in~\cite{Deser:2016tgn}, it is possible to redefine a multi-derivative Lagrangian in terms of products of 4-index Riemann tensors only. This is achieved by a gauge transformation that redefines the  metric so  as to include terms that depend linearly on the Ricci scalar and tensor. Such a manifestly gauge-invariant  action can then be used as a means for identifying the one-particle-irreducible (1PI) scattering amplitudes.

In a previous treatment~\cite{Shawa:2017exh}, we calculated a certain class of graviton multi-point  scattering amplitudes for  higher-derivative theories in the bulk of an (asymptotically) AdS spacetime. This was achieved by restricting considerations to what was originally described in~\cite{Brustein:2012he} as the ``high-momentum" regime. A  detailed description of this kinematic regime will be presented  in Subsection~1.2.
 Note that, in the original discussion  on the high-momentum regime, only Lovelock extensions to Einstein's theory were considered. However, our work is different (both in~\cite{Shawa:2017exh} and here) since it considers a generic series of products of Riemann tensors. As a consequence, our analysis allows for the possibility of two derivatives acting on a single graviton in the scattering amplitude. This possibility does not occur for Lovelock gravity; in part, because of the on-shell constraint which is imposed by the Lovelock field equations (any of which have a maximum of two derivatives per term~\cite{Lovelock:1971yv}).

The central theme of the current analysis is to use the duality to translate these earlier findings into the language of the gauge theory. The motivation is that our new results might then provide a means for identifying, experimentally,
the 
dual gravitational theory for   a  strongly coupled fluid like the 
quark--gluon plasma (QGP). A crucial step in this program is the mapping of the previously calculated  graviton scattering amplitudes to their dually related 
stress-tensor correlation functions; {\em cf}, Section~2. This entails taking the boundary limit of the amplitudes and then applying  a procedure of holographic renormalization~\cite{deBoer:1999tgo,Skenderis:2002wp}.
We also need to  determine the  connected correlation functions, as their 1PI counterparts are insufficient for the purpose of analyzing experimental data from the gauge theory. This process is detailed at length in Section~3, followed by
 a briefer discussion on experimental considerations in Section~4 and
an overview in Section~5.

Our results and methods could, at least in principle, be tested experimentally in high-energy, particle-collider laboratories such as the Large Hadron Collider (LHC) at CERN or the Relativistic Heavy Ion Collider (RHIC) at Brookhaven. In laboratories such as these, the QGP is produced and then scrutinized  via the collision of heavy ions. For a detailed discussion on how the resultant data could be used to test our findings, see~\cite{Brustein:2012dg}.

\subsection{Some notations and conventions}\label{conventions}
In this paper, we adopt the conventions of~\cite{tHooft:1974toh} in which the metric determinant and the contravariant metric are perturbatively expanded with respect to the graviton as follows:
\begin{align}
\sqrt{-g} \;\to\; \sqrt{-g}\bigg[  1 + \frac{1}{2} h_\mu^\mu - \frac{1}{2^2} h_\mu^\nu h_\nu^\mu + \mathcal{O}(h^3) \bigg]\;,
\end{align} 
\begin{align}
g^{\mu\nu} \;\to\;  g^{\mu\nu} - h^{\mu\nu}  + h^{\mu}_\rho h^{\nu\rho} + \mathcal{O}(h^3)\;. 
\end{align}
The expansions of current interest are of the forms
\begin{align}
	g^{xx}\; \to\; g^{xx} + h^x_yh^{yx} + (h^x_yh^{yx})^2 + \cdots + (h^x_yh^{yx})^p \;,
\label{mdet}
\end{align}
\begin{align}
	\sqrt{-g}\; \to\;  \sqrt{-g}\left[ 1 - \frac{1}{2}h^x_yh^y_x - \frac{1}{2^2 2!}(h^x_yh^y_x)^2 + \cdots + \Theta(p)(h^x_yh^y_x)^p \right]\;,
\label{hooft4}
\end{align}
where $x$ and $y$ are interchangeable and
\begin{align}
	\Theta(p)\; = \; -\frac{\Gamma[p-\frac{1}{2}]}{2\sqrt{\pi}p!}\;, 
\;\;\;\text{   for } p \in \mathbb{Z}\;. \label{Theta}
\end{align}

A product of $k$ Riemann tensors will either be written as Riem$^k$ or $R^k$. It is implied that such a term includes all the possible invariants that can be constructed using four-index Riemann tensors with no self-contracting indices. As an example, for Riem$^3$,
\begin{align}
\text{Riem}^3 \equiv R^3\; =\; c_1 \tensor{R}{_{ab}^{cd}}\tensor{R}{^{mn}_{cd}}\tensor{R}{^{ab}_{mn}} + c_2 \tensor{R}{_{abcd}}\tensor{R}{_{mn}^{ad}}\tensor{R}{^{mncb}}\;,
\end{align}  
where $c_1$ and $c_2$ are model-dependent constants. Later on, a $\sum_k {\rm Riem}^k$ Lagrangian will be expanded about the small parameter $\epsilon= \frac{\ell_p^2}{L^2}$, where $\ell_p$ is the Planck length and $L$ is the AdS radius of curvature.   

When constructing connected functions, we will use the inner-product symbol
 to denote a convolution of two correlation functions. For example, a convolution of a pair of four-point functions will be denoted as
\begin{align}
\avg{h^4} \cdot \avg{h^4}\;.
\end{align} 
Furthermore, when using the term ``connected function", we are excluding those that are 1PI unless stated otherwise. Finally, even when it is not stated explicitly, the high-momentum  regime (as defined below) is always in effect.

\subsection{The high-momentum regime}\label{High-momentum-regime}

In this subsection, we will recall the arena in which our calculations are performed. 

The bulk portion of the analysis is carried out  in a 5-dimensional  AdS, black brane background because this geometry is believed to provide a dual description of the 
QGP~\cite{Witten:1998zw}. However, our  current focus  is mostly on the AdS boundary, where the line element asymptotically limits to
\begin{align}\label{AAdS}
\lim_{r \to \infty} ds^2 \; = \;  \frac{r^2}{L^2}(- dt^2 + dx^2 + dy^2 + dz^2 ) + \frac{L^2}{r^2} dr^2 \;.
\end{align}
Here,  $r$ is the AdS radial coordinate  and $x$, $y$, $z$ run parallel to the brane. 

Recall that gravitons $h_{\mu\nu}$ are small perturbations of the background metric $g_{\mu\nu} \to g_{\mu\nu} + h_{\mu\nu}$. If it is assumed (without loss of generality) that they propagate along the $z$ direction, then an appropriate ansatz would be 
\begin{align}\label{def-graviton}
 h_{\mu\nu}\; = \;\phi (r) e^{i(\omega t - k z)}\;,
\end{align}  
where $\omega$ is the angular frequency, $k$ is the wavenumber and $\phi (r)$ is a finite and continuous function that is subject to the Dirichlet condition at 
the outer  boundary and the condition of a strictly incoming wave at the horizon of the black brane~\cite{Kovtun:2005ev}. The function $\phi(r)$ is of the form
\begin{align}
\phi(r)\; =\; (-g_{tt})^{\frac{i\omega}{2\pi T}}\psi(r)\;,
\end{align}
where  $\psi (r)$ is finite and continuous near the horizon of the black brane and $T$ is the Hawking temperature.

As previously stated, the process of holographic renormalization is necessary when it comes to relating graviton scattering amplitudes to stress-tensor correlation functions in the gauge theory. As shown in~\cite{Brustein:2012he} and as elaborated on later, this process implies that radial derivatives acting on gravitons need not be considered, as their contributions to the amplitudes do not survive the renormalization process. As a result, we only need to consider 
differentiations with respect to  $t$ and $z$ as far as gravitons are concerned.

As also mentioned earlier, all our calculations take place in what is known as the high-momentum regime. This regime requires that graviton frequencies and wavenumbers  are much larger than any background contribution. Since background derivatives are of the form $\nabla_r\nabla_r \sim \frac{1}{L^2}$, while $t$ and $z$ derivatives acting on  gravitons go as $\nabla_t\nabla_t \sim \omega^2 $ and $\nabla_z\nabla_z \sim k^2$, this translates into $\omega^2,k^2 \gg \frac{1}{L^2}$. However, the gauge--gravity duality implies that the hydrodynamic limit should also be imposed, to some approximation, because the fluid of interest on the gauge-theory side is supposed to be of a high temperature~\cite{Witten:1998zw}. It then follows that $k \ll T$, where $T$ is the Hawking temperature of the black brane or, equivalently, that of the fluid. Hence, the regime of interest actually is
\begin{align}
\frac{1}{L} \;\ll\; k\; \ll\; T\; .
\end{align}
Now, according to the duality, a necessary condition is that $r_h$, the value of the radial coordinate at the brane horizon, should be of the order $r_h \sim \pi TL^2 \gg L$. Meaning that the high-momentum regime is indeed consistent 
within  the framework of the  duality.

Operationally speaking, the high-momentum regime requires one to include contributions to the amplitude with the highest possible powers of $\omega$'s and $k$'s, while disregarding radial derivatives and other potential contributors (\textit{e.g.}, the cosmological-constant term). For instance, for a Lagrangian with $k$ contracted Riemann tensors (or $2k$ derivatives), one is instructed to only keep contributions carrying a factor of $\omega^p k^q$ with $p+q=2k$. We will now proceed to explain the (potential) utility of this regime in the current 
context.

For this study, the gravitational Lagrangian will be cast in the generic form,
\begin{align}\label{Lagrangian0}
\mathcal{L}\; =\;  \frac{\sqrt{-g}}{16\pi G_5}(R + a_1\epsilon L^2 R^2 + a_2\epsilon^2 L^4 R^3 + \cdots + a_k\epsilon^k L^{2k}R^{k+1} + \cdots)\;,
\end{align}
where the $a$'s are unknown constants of $\mathcal{O}(1)$ (and what we are ultimately trying to identify), $G_5$ is the five-dimensional Newton's constant and recall that $\epsilon = \frac{\ell_p^2}{L^2}$ is a perturbatively small parameter. For the meaning of $R^k$, see Subsection~\ref{conventions}. Now let us assume that, $L^2\epsilon k^2$ or $L^2\epsilon \omega^2$ are small enough so that the amplitudes eventually converge but not so small such that one is unable to detect amplitudes (and, by extension, the dual correlation functions) from higher-order $\epsilon$ terms when considering experimental data. If this is indeed the case, then the high-momentum regime allows for the possibility of experimentally distinguishing between different orders of the perturbative expansion.

In addition, we work in the radial gauge, $h_{rs} = 0$ for any $s$. In which case, the graviton splits into three types of modes: scalar $\{h_{xx},h_{yy},h_{tt},h_{zz} \}$, vector $\{ h_{xt},h_{xz},h_{yt},h_{yz} \}$ and tensor $\{ h_{xy} \}$. Scalar modes can be disregarded since they would need to be coupled to a source. The point is that, at each order in the expansion, sources would inevitably add a factor of at least one $\epsilon$, which then violates the rules of the high-momentum regime (the rule now being that any factor of $\epsilon^{k}$ should be accompanied by a factor of $\omega^p k^q$ with $p+q=2k+2$). Vector modes either could  be gauged away or, like scalar modes, would require a source for gauge-invariant combinations. This leaves us with covariant pairs of tensor modes as the only relevant types for the high-momentum regime~\cite{tHooft:1974toh}. Thus, for current purposes, all scattering amplitudes will be $2n$-point functions; {\em cf},
Eqs.~(\ref{mdet}) and~(\ref{hooft4}).

\section{Graviton amplitudes at the boundary}

In~\cite{Shawa:2017exh}, we calculated the 1PI graviton scattering amplitudes for the boundary-limit of the bulk AdS theory. But recall that, in order to make  contact with experiment, what is really needed are the connected multi-point amplitudes in the same limit. Before considering the connected  amplitudes, we will first review the methods and results of the earlier treatment. 

Let us begin here with what was referred to as  ``basis amplitudes" in the previous paper. These are scattering amplitudes  for which all participating gravitons carry at least one derivative. As in~\cite{Shawa:2017exh}, a basis amplitude with $2q$ gravitons will be denoted as $\avg{h^{2q}}_{2q}$. In this notation, the superscript denotes the total number of gravitons, while the subscript indicates the number carrying at least one derivative. (As for the total  number of 
derivatives, this  is determined by the order of the  expansion and has been  left implicit.)  Given that the high-momentum regime is in effect,  the basis amplitudes can be used to construct  all other relevant $2n$-point amplitudes, 
$\avg{h^{2n}}_{2q}$ with 
$n > q$. This is done  
 through the addition of undifferentiated gravitons, either from the contravariant metric tensors or from the metric  determinant. One can then determine the $2n$-point graviton amplitudes by summing over  all such constructions.

In the case of Riem$^3$ gravity, for instance, there are two basis amplitudes ---  a four and six-point basis amplitude --- that schematically take the form
\begin{align}
\avg{h^4}_4  &\;\sim\; c_1 \sqrt{-g}(g^{xx}g^{yy})^2 \nabla_m \nabla^n h_{xy} \nabla_n h_{xy} \nabla^l h_{xy} \nabla_l \nabla^m h_{xy}\;, \\
\avg{h^6}_6  &\;\sim\; c_2 \sqrt{-g}(g^{xx}g^{yy})^3 \nabla_m h_{xy} \nabla^m h_{xy} \nabla_n h_{xy}\nabla^n h_{xy}\nabla_l h_{xy}\nabla^l h_{xy}\;,
\end{align}
where $c_1$ and $c_2$ are model-dependent constants (\textit{i.e.}, they depend on the $a$'s in~\cref{Lagrangian0}) and $l,m,n \in \{t,z \}$.

In the   $r \to \infty$ limit, these basis amplitudes can be 
 calculated precisely,
\begin{align}
\lim_{r \to \infty}\avg{h^4}_4 &\;=\; -6c_1 (\frac{L}{r})^{11} C_{12}C_{34}C_{14} h_{xy}^{(1)}h_{xy}^{(2)}h_{xy}^{(3)}h_{xy}^{(4)}\;,  \label{R3BA1} \\
\lim_{r \to \infty}\avg{h^6}_6 &\;=\; -\frac{3}{2}c_2 (\frac{L}{r})^{15} C_{12}C_{34}C_{56} h_{xy}^{(1)}h_{xy}^{(2)}h_{xy}^{(3)}h_{xy}^{(4)}h_{xy}^{(5)}h_{xy}^{(6)}\;, \label{R3BA2}
\end{align}  
where $C_{ij} = (\frac{r}{L})^2(k_i g^{zz} k_j - \omega_i g^{tt} \omega_j)$ is a short-hand notation for contractions of products of momenta for the  gravitons $h_{xy}^{(i)}$ and $h_{xy}^{(j)}$. Note that products of $C_{ij}$'s should be symmetrized to account for all possible distinct arrangements. For example, the product of contractions in~\cref{R3BA1} actually implies
\begin{align}
C_{12}C_{34}C_{14} \;\to\; & C_{(12}C_{34}C_{14)} + C_{(12}C_{34}C_{12)} + C_{(12}C_{34}C_{13)} \notag \\ &+ C_{(12}C_{34}C_{23)} + C_{(12}C_{34}C_{24)} + C_{(12}C_{34}C_{34)}\;, 
\end{align}
where terms with repeating numbers indicate the presence of two derivatives acting on a graviton. In addition, the field equation $\Box h_{xy} = 0 + \mathcal{O}(\epsilon)$ has been used to eliminate any appearance of 
$C_{ii}\propto \Box h^{(i)}_{xy}$. 

For the Riem$^3$ theory, we can now express the complete  $6$-point amplitude at the boundary, which is a sum of all such contributions, as
\begin{align}
\lim_{r \to \infty}\avg{h^6} &\;=\; \lim_{r \to \infty}\avg{h^6}_4 + \lim_{r \to \infty} \avg{h^6}_6 \\
&\;=\; -315 c_1  (\frac{L}{r})^{15} \; \; C_{12}C_{34}C_{14} \; \; h_{xy}^{(1)}h_{xy}^{(2)}h_{xy}^{(3)}h_{xy}^{(4)}h^{(5)}_{xy}h^{(6)}_{xy}  \notag \\
&- \frac{3}{2} c_2 (\frac{L}{r})^{15} \; \; C_{12}C_{34}C_{56} \; \; h_{xy}^{(1)}h_{xy}^{(2)}h_{xy}^{(3)}h_{xy}^{(4)}h_{xy}^{(5)}h_{xy}^{(6)}\;. \label{R3CA}
\end{align}
More generally, the $2n$-point amplitude for this theory has a  boundary form
of
\begin{align}
\lim_{r \to \infty} \avg{h^{2n}}_{\text{Riem}^3} &= \lim_{r \to \infty} \avg{h^{2n}}_4  + \lim_{r \to \infty} \avg{h^{2n}}_6 \\
&=  - \bigg(\frac{L}{r} \bigg)^{4n+3} \bigg[ 6c_1\begin{pmatrix} 2n \\ 4 \end{pmatrix}\sum_{p=0}^{n-2} \begin{pmatrix} n+1-p \\ 3 \end{pmatrix} \Theta(p) \; \; C_{12}C_{34}C_{14}  \notag \\ &\ \ +\frac{3}{2}c_2\begin{pmatrix} 2n \\ 6 \end{pmatrix}\sum_{p=0}^{n-3} \begin{pmatrix} n+2-p \\ 5 \end{pmatrix} \Theta(p) \; \; C_{12}C_{34}C_{56} \; \; \bigg] \Pi_{k=1}^{2n} h_{xy}^{(k)}\;,
\end{align} 
with $\Theta(p)$ as defined in Eq.~(\ref{Theta}).

In similar fashion, one can construct the boundary $2n$-point amplitude for the Riem$^4$ theory, while taking into account that it has three basis amplitudes, 
and  arrive at
\begin{align}
\lim_{r \to \infty} \avg{h^{2n}}_{\text{Riem}^4} &\;=\; \lim_{r \to \infty} \avg{h^{2n}}_4  + \lim_{r \to \infty} \avg{h^{2n}}_6 +  \lim_{r \to \infty} \avg{h^{2n}}_8 \\ 
&\;=\; \bigg(\frac{L}{r} \bigg)^{4n+5}\bigg[ b_1  \begin{pmatrix} 2n \\ 4 \end{pmatrix}\sum_{p=0}^{n-2} \begin{pmatrix} n+1-p \\ 3 \end{pmatrix} \Theta(p) \; \; C_{12}C_{23}C_{34}C_{14} \notag \\ 
&\; \; + b_2  \begin{pmatrix} 2n \\ 6 \end{pmatrix}\sum_{p=0}^{n-3} \begin{pmatrix} n+2-p \\ 5 \end{pmatrix} \Theta(p) \; \; C_{12}C_{23}C_{34}C_{56}  \notag \\
&\; \; + b_3 \begin{pmatrix} 2n \\ 8 \end{pmatrix}\sum_{p=0}^{n-4} \begin{pmatrix} n+3-p \\ 7 \end{pmatrix} \Theta(p) \; \; C_{12}C_{34}C_{56}C_{78} \bigg] \Pi_{k=1}^{2n} h_{xy}^{(k)}\;,\label{R4CA}
\end{align}
where $b_1$, $b_2$ and $b_3$ are model-dependent constants.

Moreover, as shown  in~\cite{Shawa:2017exh}, it is possible to obtain the boundary limit of the $2n$-point amplitude for an arbitrary Riem$^k$ theory. (Note that the number of basis amplitudes depends on both the size and parity of $k$.) To demonstrate  how this works, let us consider a Riem$^k$ theory with a basis amplitude given by $\avg{h^{2q}}_{2q}$. The determinant $\sqrt{-g}$ contributes a factor~$(\frac{L^2}{r^2})^{-\frac{3}{2}}$, the contravariant metrics 
$(g^{xx}g^{yy})^q$ contribute a factor of~$(\frac{L^2}{r^2})^{2q}$ and the $k$ 
contractions of momenta contribute a factor~$(\frac{L^2}{r^2})^{k}$. The inclusion of undifferentiated gravitons adds invariant pairs $h_x^y h_y^x$, which can also be written as $g^{xx}g^{yy} h_{xy}h_{xy}$. And so the undifferentiated gravitons contribute an overall factor of~$(\frac{L^2}{r^2})^{2n-2q}$. In total, there is a contribution of~$(\frac{L}{r})^{4n+2k-3}$. We can now express the boundary limit of the 1PI, $2n$-point amplitude for the Riem$^k$ theory in the schematic form
\begin{align}\label{basistor}
\lim_{r \to \infty} \avg{h^{2n}}_{\text{Riem}^k} &\;= \; \bigg(\frac{L}{r} \bigg)^{4n + 2k - 3} \; \sum_{q}  \avg{h^{2n}}_{2q}\;,
\end{align}      
where the summation implies adding the contributions from all relevant basis amplitudes (as these all make dimensionally equivalent contributions).

One can now apply the process of holographic renormalization via the following steps:
\begin{enumerate}
\item Take the bulk amplitudes to the boundary (as already done). 
\item Multiply the bulk-to-boundary amplitudes by a conformal factor $\Omega^v$, where $v$ is the sum of the conformal dimensions of the operators, which in this case are the gravitons and derivatives. The metric determinant is also known to make an effective contribution to the power of $\Omega$~\cite{Bianchi:2001kw}.
\item Remove any subsequent divergences using boundary-localized counter terms~\cite{deBoer:2000cz}. However, the high-momentum regime renders this step irrelevant. Had we considered radial derivatives on gravitons, then divergent terms would be present and would have been cancelled off by counter terms. On the contrary, the terms with $z$ and $t$ derivatives are already finite. 
\end{enumerate}

For Step 2, the conformal factor is deduced from the asymptotic form of the AdS metric as given by~\cref{AAdS}, $\Omega = \frac{r}{L}$~\cite{deBoer:1999tgo}. 
Each derivative has of course a conformal dimension of 1, while each  graviton 
has a dimension of 2. The conformal dimension of the graviton follows from the boundary behavior of the metric. In particular, for a rescaling $r \to \gamma r$, then $g_{\mu\nu} \to \gamma^{-2} g_{\mu\nu}$. The square root of the metric determinant contributes an effective conformal dimension of $-3$. Hence, for a $2n$-point amplitude at the order of  $\epsilon^{k-1}$ (or $2k$ derivatives), the power of the conformal factor is given by $\Omega^{(2n\times 2) +  2k - 3} = (\frac{r}{L})^{4n + 2k - 3}$. As  evident from~\cref{basistor}, which carries the same factor of $\epsilon^{k-1}$, this process eliminates any $r$ dependence in the gauge-theory duals of the amplitudes.

\begin{figure}[H]
\centering
	\begin{subfigure}[t]{0.2\textwidth}
	\includegraphics[scale=0.75]{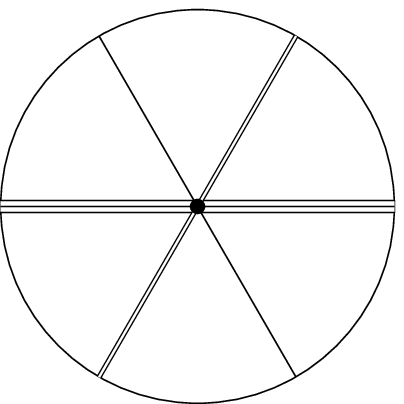}
	\caption{}
	\label{WdR31PIa}
	\end{subfigure}
~~~~~~~~~
	\begin{subfigure}[t]{0.2\textwidth}
	\includegraphics[scale=0.75]{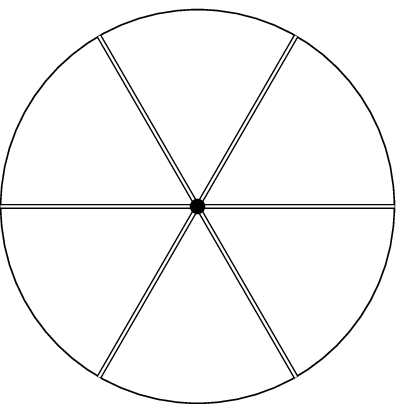}
	\caption{}
	\label{WdR31PIb}
	\end{subfigure}
~~~~~~~~
	\begin{subfigure}[t]{0.2\textwidth}
	\includegraphics[scale=0.75]{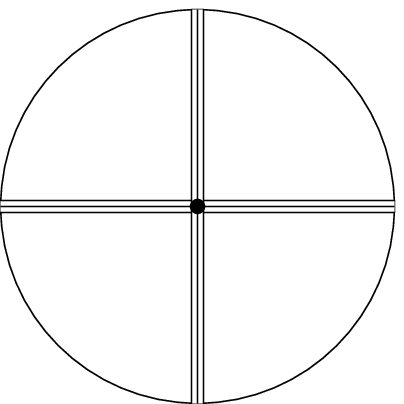}
	\caption{}
	\label{r41pixx}
	\end{subfigure}
~~~~~~~~
	\begin{subfigure}[t]{0.2\textwidth}
	\includegraphics[scale=0.75]{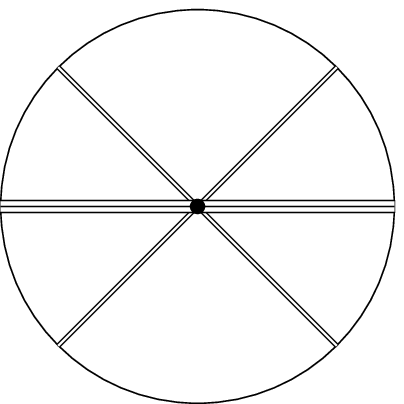}		
	\caption{}
	\label{R461PIa}
	\end{subfigure}
~~~~~~~~
	\begin{subfigure}[t]{0.2\textwidth}
	\includegraphics[scale=0.75]{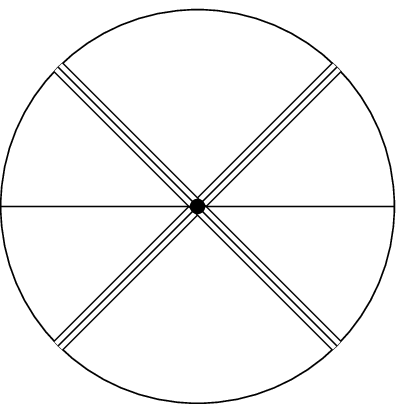}	
	\caption{}
	\label{R461PIb}
	\end{subfigure}
~~~~~~~~
	\begin{subfigure}[t]{0.2\textwidth}
	\includegraphics[scale=0.75]{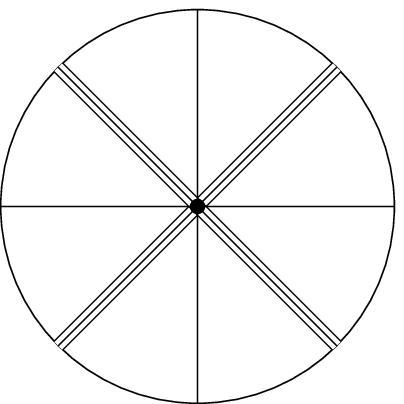}
	\caption{}
	\label{R481PIa}
	\end{subfigure}
~~~~~~~~
	\begin{subfigure}[t]{0.2\textwidth}
	\includegraphics[scale=0.75]{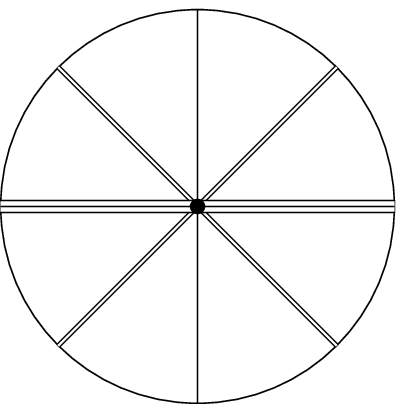}
	\caption{}
	\label{R481PIb}
	\end{subfigure}
~~~~~~~~~
	\begin{subfigure}[t]{0.2\textwidth}
	\includegraphics[scale=0.75]{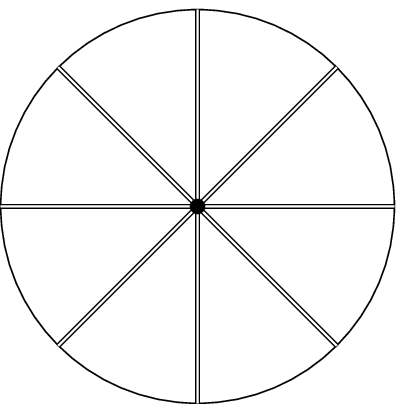}
	\caption{}
	\label{R481PIc}
	\end{subfigure}

\caption{One-particle-irreducible Witten diagrams at orders $\epsilon^2$ and $\epsilon^3$. Single lines denote undifferentiated gravitons, double lines denote gravitons carrying a single derivative and triple lines denote gravitons carrying two derivatives.~Figures~\ref{WdR31PIa} and \ref{WdR31PIb} are $\epsilon^2$-order $6$-point functions. Figures~\ref{r41pixx} to \ref{R481PIc} depict various 
$\epsilon^3$-order functions.
}
\label{1PIsDiagrams}
\end{figure}

The Witten diagrams~\cite{Witten} for some of the renormalized 1PI amplitudes are shown in~\cref{1PIsDiagrams}.

\section{Connected functions}
We have, so far, been calculating  multi-point functions by expanding the
gravitational Lagrangian. By definition, all the multi-point functions under consideration are really derived from a generating functional of the schematic form
\begin{align}\label{genFun}
e^{i W[J]} &\;=\; \int \mathcal{D}[h] e^{i\int d^5x[ \sqrt{-g} (a_0 R[h] + a_1 \epsilon R^2[h] + a_2 \epsilon^2 R^3[h] + a_3 \epsilon^3 R^4[h] + \cdots + a_k \epsilon^k R^{k+1} + \cdots) + Jh]}\;, 
\end{align}
where $J$ is the source term. To obtain connected functions, one is instructed to vary the generating functional with respect to the source,
\begin{align}
\avg{h_1h_2\cdots h_{2n}} &\;=\; (-i)^{2n+1}\bigg[ \frac{\delta^n W}{\delta J_1 \cdots \delta J_{2n}} \bigg]_{J=0}\;,
\end{align}
where the subscripts on the $h$'s and  $J$'s denote points in spacetime. 

At each order in $\epsilon$, there are both single-vertex  and multi-vertex diagrams to consider. Single-vertex diagrams, also known as 1PI functions, can be read directly off the Lagrangian. On the other hand, multi-vertex diagrams can be constructed by convolving lower-order 1PI diagrams. At the  order of  $\epsilon^k$, a two-vertex, $2n$-point function which is formed by convolving a $2p$- and a $2q$-point function can be schematically expressed as
\begin{align}
\avg{h^{2n}}^{\epsilon^k} &\;=\; \avg{h^{2p}}^{\epsilon^i}\cdot\avg{h^{2q}}^{\epsilon^j}\;,
\end{align} 
as long as $2n = 2p + 2q - 2$ and $k = i + j$. Notice that this process consumes a pair of gravitons and a pair of derivatives through the propagator connecting the two vertices.  However, as discussed in~\cite{Brustein:2012pq}, a convolution of any  pair of 1PI  diagrams at the orders of  $\epsilon^k$  and  $\epsilon^0$ always returns a 1PI diagram of order$\epsilon^k$. This is because the $\epsilon^0$-order theory is a two-derivative theory and two derivatives would be used up in the convolution process. The end result is a $2k+2$ derivative structure, which is consistent with an $\epsilon^k$-order theory. But, for higher powers of $\epsilon$, one can rather expect a wide variety of non-vanishing connected functions. 

The high-momentum regime sets a  limit on the  number of theories that can contribute to any given  amplitude, as we now explain: For a given  set of basis amplitudes, there is always a ``smallest" amplitude; that for  which the maximal number of  participating gravitons are twice differentiated. For instance, 
a ${\rm  Riem}^k$ theory has a smallest amplitude of $\avg{h^k}$ if $k$ is even and $\avg{h^{k+1}}$ if odd. And so, for example, a 6-point amplitude can get contributions from
${\rm Riem}^6$  in the high-momentum regime but not from ${\rm Riem}^7$ (nor
for  any theory of even higher order).

The analysis to follow will only include tree-level amplitudes as 
$N$ is presumed to be very large (albeit finite). Consequently, it can be argued that, regardless of the order in $\epsilon$, all connected $4$-point amplitudes are 1PI in the high-momentum regime, $\avg{h^4}_{\text{con}} = \avg{h^4}_{1PI}$. (See~\cite{Brustein:2012pq} for a detailed explanation.) Higher-point functions will, however, include a mixture of 1PI and connected amplitudes.

\subsection{Connected diagrams at order $\epsilon^2$}\label{e2diagrams}

At the order of $\epsilon^2$, one can use the generating functional in~\cref{genFun} to show that, whereas the 1PI amplitudes are generated from the Riem$^3$ term,
\begin{align}
\avg{h^{2n}}_{\text{1PI}}^{\epsilon^2} &\;=\; a_2\epsilon^2 \int d^5x \sqrt{-g}R^3_{\mathcal{O}(h^{2n})}\;,
\end{align}
the connected amplitudes are generated by expanding the exponential in~\cref{genFun} and then collecting all the $\epsilon^2$-order terms after the source has been varied. Schematically, such  a connected function might take the form 
\begin{align}\label{R3exp1}
\avg{h^{2n}}_{\text{con}}^{\epsilon^2}  &\;=\; \frac{a_1^2}{2!} \epsilon^2 \int d^5x \int d^5y \sqrt{-g(x)} \sqrt{-g(y)} (R^2(x) \cdot R^2(y))_{\mathcal{O}(h^{2n})}\;, 
\end{align}
or, perhaps,
\begin{align}\label{R3exp2} 
\avg{h^{2n}}_{\text{con}}^{\epsilon^2}   &\;=\;  \frac{a_1^2}{2!} \epsilon^2 \int d^5x \int d^5y \int d^5z \sqrt{-g(z)} \sqrt{-g(x)} \sqrt{-g(y)} (R^2(x) \cdot R^2(y) \cdot R(z))_{\mathcal{O}(h^{2n})}\;.
\end{align}  
In this context, a  centered dot  on the right-hand side represents  a propagator connecting a pair of vertices.

\begin{figure}[H]
\centering
\includegraphics[scale=0.75]{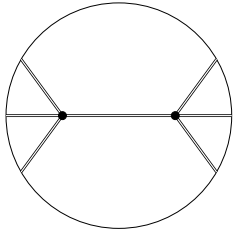}
	\caption{Witten diagram for a connected $6$-point amplitude at order $\epsilon^2$.}
	\label{WdR31PRc}
\end{figure}

The  simplest non-trivial, non-vanishing  convolution would  be  a pair of  $\epsilon$-order $4$-point amplitudes convolved to obtain a $6$-point function of order 
$\epsilon^2$. This calculation  has already  been performed in~\cite{Brustein:2012pq} and led to an  outcome of
\begin{align}
\avg{h^6}_{con}^{\epsilon^2} &\;=\; \avg{h^4}^{\epsilon}_{1PI}\cdot \avg{h^4}^{\epsilon}_{1PI} \notag \\ 
&\;=\; -\frac{1}{3}\int dr \int d^4x \sqrt{-g(r,x)} \nabla_e h_{ab}(x) \nabla^e h^{ab}(x) \nabla_f h_{mn}(x) \nabla^f h^{mn}(x) \notag \\  & \; \;\;\;\;\;\;\; \; \;\;\;\;\;\;\; \times \nabla_g h_{pq}(x) \nabla^g h^{pq}(x)\;. 
\end{align} 
The corresponding diagram is shown in~\cref{WdR31PRc}.

Without (much) loss of generality, we will illustrate the general procedure by presenting a detailed calculation of a two-vertex, connected $8$-point 
function at the order of  $\epsilon^2$. Diagrams with an even larger number of vertices can be calculated using the same basic methods. The case of current interest entails the convolution of a $6$- and a $4$-point function. The first of what will be three main steps is to determine the contribution of this
form of  connected function \textit{relative} to a 1PI $8$-point function as far as the Feynman combinatorics is concerned (as the ``strength" of each 1PI function is already known or can be readily worked out).  This theory-independent ratio of amplitudes will be generally  referred to as the  ``relative Feynman weight" and is 
determined to be $28$ for the current case. (The methodology  
is  explained in  \cite{Brustein:2012pq}.) 

\begin{figure}[H]
~ ~ ~ ~ ~ ~ ~ ~ ~ ~	
\begin{subfigure}[t]{0.2\textwidth}
		\includegraphics[scale=0.75]{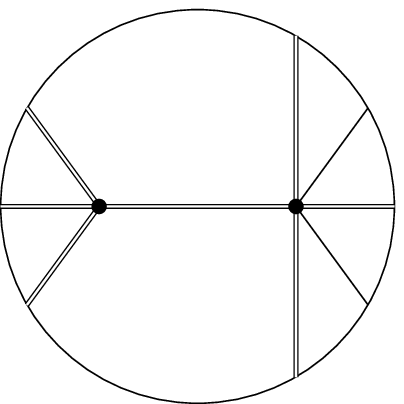}
		\caption{}
		\label{R38e}
\end{subfigure}
~ ~ ~ ~ ~ ~ ~ ~ ~ ~	
\begin{subfigure}[t]{0.2\textwidth}
		\includegraphics[scale=0.75]{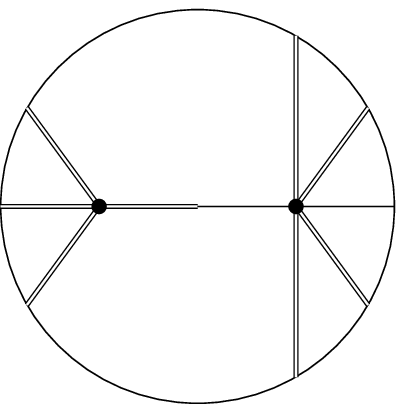}
		\caption{}
		\label{R38f}
\end{subfigure}
\caption{Connected $8$-point diagrams at order $\epsilon^2$. Figure~\ref{R38e} depicts a propagator with both gravitons differentiated and  Fig.~\ref{R38f} shows an undifferentiated graviton in the propagator. }
\label{R3864diagrams}
\end{figure}

The second step is the actual convolving of the amplitudes.
Since using an $\epsilon^0$-order function in a two-vertex convolution would 
give back
a 1PI amplitude, we need to convolve a pair of $\epsilon$-order functions; one and only one of which (the 6-point amplitude) will necessarily contain 
a pair of undifferentiated gravitons. (Two derivatives cannot act on a graviton in this case because any $\epsilon^1$-order Lagrangian can be gauge transformed into a Lovelock theory; namely,   Gauss--Bonnet gravity.) Then, with respect to the propagator, there are two possible configurations to consider: The propagator either has two differentiated gravitons or it carries only a single differentiated graviton, as depicted in~\cref{R38e,R38f} respectively. 
(There is also an equal chance that the 6-point function appears  on the ``left"- or the ``right"-hand side of both types of  Witten diagrams.)
Our analysis for the second step begins with the former ``both-differentiated'' case. 

\emph{Case (a)}: The connected function in \cref{R38e} can be written as 
\begin{align}\label{startcalc1}
\avg{h^8}^{\epsilon^2, (a)}_{con} &\; =\; \int dr \int d^4x \int d\bar{r} \int d^4\bar{x} \; \; h(r,x) h(r,x) \nabla_a h(r,x) \nabla^a h(r,x) \nabla_b h(r,x)  \notag \\ & \; \; \; \; \; \;\; \; \; \; \; \; \; \; \; \; \; \; \times  \avg{\nabla^b h(r,x) \bar{\nabla}_c h(\bar{r},\bar{x})} \bar{\nabla}^c h(\bar{r},\bar{x}) \bar{\nabla}_d h(\bar{r},\bar{x}) \bar{\nabla}^d h(\bar{r},\bar{x})\;,
\end{align}
where  the  ``expectation value'' on the right depicts the propagator and  the (implicit) tensor structure of the gravitons will be dealt with later. For now, the objective  is to simplify this expression by eliminating the propagator. We start by applying integration by parts to get
\begin{align}
\avg{h^8}^{\epsilon^2, (a)}_{con} &\;=\; - \int dr \int d^4x \int d\bar{r} \int d^4\bar{x} \; \nabla^b \bigg[ h(r,x)h(r,x) \nabla_a h(r,x) \nabla^a h(r,x) \nabla_b h(r,x) \bigg]  \notag \\ & \; \; \; \; \; \;\; \; \; \; \; \; \; \; \; \; \; \; \times \avg{h(r,x)\bar{\nabla}_ch(\bar{r},\bar{x})} \bar{\nabla}^c h(\bar{r},\bar{x}) \bar{\nabla}_d h(\bar{r},\bar{x}) \bar{\nabla}^d h(\bar{r},\bar{x})\;.
\end{align}
The product rule and the indistinguishability of the gravitons are then
applied to give the following:
\begin{align}\label{some-intermediate-calculation}
\avg{h^8}^{\epsilon^2, (a)}_{con} &\;=\; - 4 \int dr \int d^4x \int d\bar{r} \int d^4\bar{x} \; \; \; h(r,x) \nabla^b h(r,x) \nabla_a h(r,x) \nabla^a h(r,x) \nabla_b h(r,x)  \notag \\ & \; \; \; \; \; \;\; \; \; \; \; \; \; \; \; \; \; \; \times  \avg{h(r,x)\bar{\nabla}_ch(\bar{r},\bar{x})} \bar{\nabla}^c h(\bar{r},\bar{x}) \bar{\nabla}_d h(\bar{r},\bar{x}) \bar{\nabla}^d h(\bar{r},\bar{x})\;.
\end{align}

Now moving  to the $(\bar{r},\bar{x})$-side of the calculation, we apply the inverse of the product rule to obtain 
\begin{align}\label{midcalc1}
\avg{h^8}^{\epsilon^2, (a)}_{con} &\;=\;  -\frac{4}{3} \int dr \int d^4x \int d\bar{r} \int d^4\bar{x} \; \; \; h(r,x)\nabla^b h(r,x) \nabla_a h(r,x) \nabla^a h(r,x) \nabla_b h(r,x)   \notag \\ & \; \; \; \; \; \;\; \; \; \; \; \; \; \; \; \; \; \; \times \bar{\nabla}_c\avg{h(r,x)h(\bar{r},\bar{x})} \bar{\nabla}^c \bigg[ h(\bar{r},\bar{x}) \bar{\nabla}_d h(\bar{r},\bar{x}) \bar{\nabla}^d h(\bar{r},\bar{x})\bigg]\;,
\end{align}
which  is then followed by an integration by parts,
\begin{align}\label{midcalc2.0}
\avg{h^8}^{\epsilon^2, (a)}_{con} &\;=\;  \frac{4}{3} \int dr \int d^4x \int d\bar{r} \int d^4\bar{x} \; \; \; h(r,x)\nabla^b h(r,x) \nabla_a h(r,x) \nabla^a h(r,x) \nabla_b h(r,x)  \notag \\ & \; \; \; \; \; \;\; \; \; \; \; \; \; \; \; \; \; \; \times  \bar{\nabla}^c\bar{\nabla}_c\avg{h(r,x)h(\bar{r},\bar{x})}   h(\bar{r},\bar{x}) \bar{\nabla}_d h(\bar{r},\bar{x}) \bar{\nabla}^d h(\bar{r},\bar{x})\;.
\end{align}

Next, the Green's function of the zeroth-order field equation,~\footnote{The 
 $\epsilon^{0}$-order field equation is used to maintain consistency with the rules of the high-momentum regime.} 
\begin{align}\label{GreensFunc}
\nabla_a \nabla^a \avg{h(r,x)h(\bar{r},\bar{x})} &\;=\; \frac{1}{\sqrt{-g(r,x)}} \delta (r-\bar{r}) \delta^{(4)}(x-\bar{x}) \notag \\
&\;=\; \frac{1}{\sqrt{-g(\bar{r},\bar{x})}} \delta (r-\bar{r}) \delta^{(4)}(x-\bar{x})\;,
\end{align}
 allows us to rewrite \Cref{midcalc2.0} as
\begin{align}\label{endcalc1}
\avg{h^8}^{\epsilon^2, (a)}_{con} &\;=\; \frac{4}{3} \int dr \int d^4x \; \; h(r,x) h(r,x) \nabla_a h(r,x) \nabla^a h(r,x) \nabla_b h(r,x)\nabla^b h(r,x) \notag \\ & \; \; \; \; \; \;\; \; \; \; \; \; \; \; \; \; \; \; \times \nabla_c h(r,x) \nabla^c h(r,x)\;,
\end{align}
a form which is consistent with theories of order  $\epsilon^2$. 

There are further considerations for  this portion  of the calculation. From~\cref{startcalc1}, we could have started with an integration by parts on the $(\bar{r},\bar{x})$ side, followed by the inverse of the product rule and then an integration by parts on the $(r,x)$ side. The resulting expression would then rather be
\begin{align}\label{midcalc2}
\avg{h^8}^{\epsilon^2, (a)}_{con}  &\;=\;  \frac{2}{5} \int dr \int d^4x \int d\bar{r} \int d^4\bar{x} \; \; \; h(r,x) h(r,x) \nabla_a h(r,x) \nabla^a h(r,x)  h(r,x)  \notag \\ & \; \; \; \; \; \;\; \; \; \; \; \; \; \; \; \; \; \; \times \nabla^b\nabla_b\avg{h(r,x)h(\bar{r},\bar{x})}  \bar{\nabla}_c h(\bar{r},\bar{x}) \bar{\nabla}^c \bar{\nabla}_d h(\bar{r},\bar{x}) \bar{\nabla}^d h(\bar{r},\bar{x})\;,
\end{align}
which  would have led to a factor of $\frac{2}{5}$ instead of  $\frac{4}{3}$. We still, however, have to consider the symmetrization of the external gravitons. For instance, in the integration over the~$(\bar{r},\bar{x})$ coordinates that led to~\cref{midcalc2.0}, it is necessary to symmetrize $3$ of the $8$ external gravitons. Similarly, in the second instance leading up to~\cref{midcalc2}, $5$ of the $8$ external gravitons must be symmetrized. And so the effective factor, after symmetrization has been implemented, becomes $\frac{3}{8}\times \frac{4}{3} + \frac{5}{8}\times \frac{2}{5} = \frac{3}{4}$. The validity of this symmetrization  procedure is further motivated in~\cite{Brustein:2012pq}.

The previously ignored tensor structure of the gravitons still has to be accounted for. Tensorially, a Riem$^2$ term can be expressed as
\begin{align}\label{first-decomposition}
\tensor{R}{_{ab}^{cd}}\tensor{R}{^{ab}_{cd}}\; =\; \mathcal{X}^{mncd}R_{abmn}\mathcal{X}^{abpq}R_{pqcd}\;,
\end{align}
where $\mathcal{X}^{abmn} = \frac{1}{2}(g^{am}g^{bn} - g^{an}g^{bm})$ is a tensor that accounts for the anti-symmetry properties of the Riemann tensor.
We can use the above expression to write one particular example of an $\epsilon$-order $4$-point function,
\begin{align}\label{1decomp1PI}
\avg{h^4}^{\epsilon}_{1PI} &\;=\; \mathcal{X}^{mncd} \nabla_e h_{an} \nabla^e h_{bm} \mathcal{X}^{abpq} \nabla_f h_{pd} \nabla^f h_{qc}\;.
\end{align}
Similarly, an example of an $\epsilon$-order $6$-point function is
\begin{align}
\avg{h^6}^{\epsilon}_{1PI} &\;=\; \mathcal{X}^{mncd} \nabla_e h_{an} \nabla^e h_{bm} \mathcal{X}^{abpq} \nabla_f h_{pd} \nabla^f h_{qc} \mathcal{Y}^{rstu}h_{ru}h_{st}\;,
\end{align}
where $\mathcal{Y}^{abcd} = \frac{1}{2}(g^{ac}g^{bd} + g^{ab}g^{cd})$ accounts for the symmetric structure of the undifferentiated gravitons.

The derivatives in these expressions can be safely dropped as they are irrelevant to the remainder of the calculation. Then, with  only the relevant tensor structure
made explicit, the convolution of a $4$- and $6$-point function of the type shown in~\cref{R38e} goes as  
\begin{align}\label{R38pfTensorStructure}
\avg{h^8}^{\epsilon^2, (a)}_{con} &\;=\;  \mathcal{X}^{klvw}h_{vj}h_{wi}\mathcal{X}^{efij} h_{le} \avg{h_{kf}h_{an}} h_{bm} \mathcal{X}^{mncd} \mathcal{X}^{abpq}  h_{pd}  h_{qc} \mathcal{Y}^{rstu}h_{ru}h_{st}\;.
\end{align} 

One can now call upon the graviton propagator in momentum space (with the momenta implied)~\cite{Zakharov:1970cc,vanDam:1970vg},~\footnote{The flat-space form of the Einstein propagator is sufficient as the cosmological constant is only relevant to scalar modes and any additional structure would add on another order of $\epsilon$.}
\begin{align}\label{GravPropagator}
\avg{h_{ab}h_{cd}} \equiv \mathcal{G}_{abcd}\; =\; \frac{1}{2} (g_{ac}g_{bd}+ g_{ad}g_{bc} - g_{ab}g_{cd})\;,
\end{align}
apply this to~\cref{R38pfTensorStructure} and, with some effort, finally obtain 
\begin{align}\label{R38pfResult}
\avg{h^8}^{\epsilon^2, (a)}_{con} &\;=\; \frac{1}{4} h_{ab}h^{ab}h_{cd}h^{cd}h_{mn}h^{mn}\mathcal{Y}^{rstu}h_{ru}h_{st}\;.
\end{align}
The simplicity of~\cref{R38pfResult} comes as a result of eliminating terms such as $h_a^a$ and $h^{ab}h_a^ch_{cb}$ as they indicate the presence of scalar graviton modes. Nonetheless, the numerical factor of $\frac{1}{4}$ is the real interest of this calculation.

\emph{Case (b)}: A similar  process can be used for the convolution
that is depicted  in~\cref{R38f}. Here, the appropriate  starting point is 
\begin{align}
\avg{h^8}^{\epsilon^2, (b)}_{con} &\;=\; \int dr \int d^4x \int d\bar{r} \int d^4\bar{x} \; \; \; h(r,x)\nabla^b h(r,x) \nabla_a h(r,x) \nabla^a h(r,x) \nabla_b h(r,x) \notag \\ & \; \; \; \; \; \;\; \; \; \; \; \; \; \; \; \; \; \; \times \avg{h(r,x)\bar{\nabla}_ch(\bar{r},\bar{x})} \bar{\nabla}^c h(\bar{r},\bar{x}) \bar{\nabla}_d h(\bar{r},\bar{x}) \bar{\nabla}^d h(\bar{r},\bar{x})\;.
\end{align}
Following the same process of symmetrizing external gravitons and moving derivatives around through the use of integration by parts and the inverse of the product rule, one ends up with the expression
\begin{align}
\avg{h^8}^{\epsilon^2, (b)}_{con} &\;=\; -\frac{3}{16}\int dr \int d^4x \; \;  h(r,x) h(r,x) \nabla_a h(r,x) \nabla^a h(r,x) \nabla_b h(r,x)\nabla^b h(r,x) 
\notag \\ & \; \; \; \; \; \;\; \; \; \; \; \; \; \; \; \; \; \; \times \nabla_d h(r,x) \nabla^d h(r,x)\;.
\end{align}
Unlike the previous case, one obtains this result irrespective
of  how the steps are implemented. Hence, the effective factor remains
at  $-\frac{3}{16}$.

As for the tensorial part of the process, this requires reducing the expression
\begin{align}
\avg{h^8}^{\epsilon^2, (b)}_{con} &\;=\;  \mathcal{X}^{klvw}h_{vj}h_{wi}\mathcal{X}^{efij} h_{le} \avg{h_{kf} h_{st}} h_{ru}  \mathcal{Y}^{rstu}h_{an} h_{bm} \mathcal{X}^{mncd} \mathcal{X}^{abpq}  h_{pd}  h_{qc}\;,
\end{align}
by way of the graviton propagator and further simplifications. The 
eventual outcome is
\begin{align}\label{R38pfResultb}
\avg{h^8}^{\epsilon^2, (b)}_{con} &\;=\; h_{ab}h^{ab} h_{cd} h^{cd} h_{mn}h^{mn}  h_{pq}  h^{pq}\; .
\end{align}

The third and final step of the  process  is to determine the net contribution from all the implicated  diagrams.  This involves  multiplying all of the numerical coefficients 
that  were obtained for each of the cases separately  and then summing over
the cases.  Importantly, the four possible diagrams,~\cref{R38e,R38f} and their respective reflections, contribute equally to the Feynman relative weight of 28 for this  two-vertex $8$-point function. 

For \emph{Case (a)} and its reflection, the net relative weight is then
\begin{align}
\frac{3}{4} \times \frac{1}{4}\; \times\;  \frac{28}{4} \times 2
\; =\; \frac{21}{8}\;.\label{positive}
\end{align}
The first factor on the left-hand side comes from convolving in position space
after symmetrization is considered. The second factor accounts for  the tensor structure of the gravitons. The third factor can be attributed to the Feynman relative weight and the fourth accounts for the reflection.

For \emph{Case (b)} and its reflection, the net relative weight is rather
\begin{align}
-\frac{3}{16} \times 1 \; \times\; \frac{28}{4} \times 2\; =\; -\frac{21}{8}\;. \label{negative}
\end{align}
Summing Eqs.~(\ref{positive}) and ~(\ref{negative}), we can conclude that there is no actual amplitude that results from this particular convolution.

The only other way to construct a connected $8$-point function would be to use  a trio of $4$-point functions. The only distinct diagrams of this form 
at the order of $\epsilon^2$
are depicted in~\cref{3vde2}\;. One then proceeds using the same methods as above.  (See Appendix~A in~\cite{Brustein:2012pq} for an example of a three-vertex 
calculation.)
\begin{figure}[H]
\centering
\begin{subfigure}[t]{0.2\textwidth}
	\includegraphics[scale=0.75]{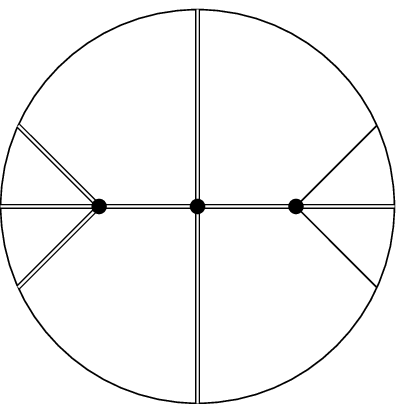}
	\caption{}
	\label{R38a}
\end{subfigure}
~~~~
\begin{subfigure}[t]{0.2\textwidth}
	\includegraphics[scale=0.75]{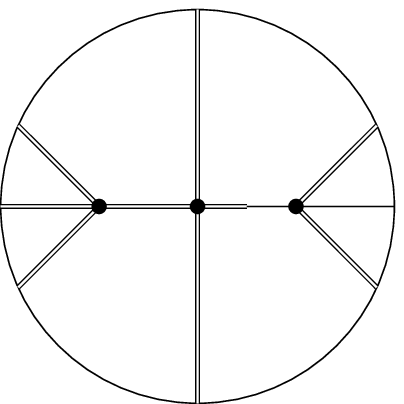}
	\caption{}
	\label{R38b}
\end{subfigure}
~~~~
\begin{subfigure}[t]{0.2\textwidth}
	\includegraphics[scale=0.75]{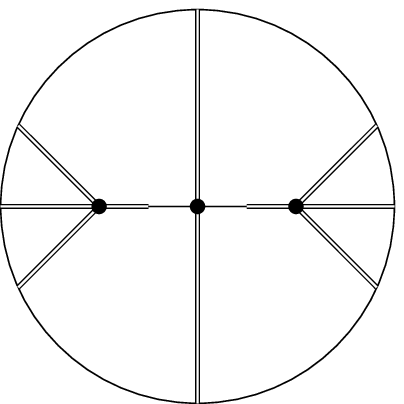}
	\caption{}
	\label{R38c}
\end{subfigure}
~~~~
\begin{subfigure}[t]{0.2\textwidth}
	\includegraphics[scale=0.75]{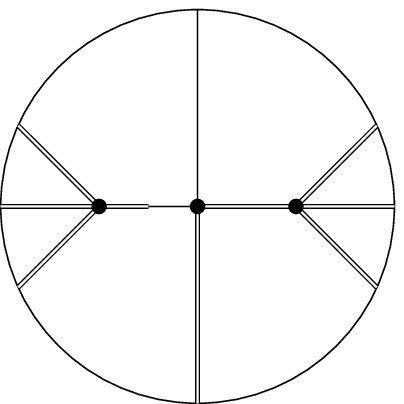}
	\caption{}
	\label{R38d}
\end{subfigure}
\caption{Three-vertex, connected $8$-point functions at order $\epsilon^2$.}
\label{3vde2}
\end{figure} 

To get higher-point functions at this same order, one can add more vertices using $\epsilon^0$-order 1PI functions or by adding as many undifferentiated gravitons as  necessary to the lower-point functions. If more vertices are indeed added, then one would have to deal with the combinatorics of all the possible positions that the $\epsilon^0$ pieces might occupy along the chain. This is a process that would inevitably introduce additional complexity to the calculation. Yet it is a straightforward task in principle.  

\subsection{Connected diagrams at $\epsilon^3$ order}

Here,  we will briefly discuss the calculation of a typical  $\epsilon^3$-order, two-vertex, connected $6$-point diagram, which  means 
convolving a pair of 1PI 4-point amplitudes at the orders of  $\epsilon$ and $\epsilon^2$
({\em i.e.,} those from Riem$^2$ and Riem$^3$ theories  respectively).
Our current motivation is to demonstrate  a subtle procedure which was previously  absent. In particular, the calculation now incorporates twice-differentiated gravitons because
  $\epsilon^2$-order gravity is not, in general, related to a Lovelock theory by a gauge transformation.  
Higher-point functions, even at higher orders in $\epsilon$, can be calculated using the same basic reasoning as in the examples provided. 
 
\begin{figure}[H]
\centering
	\begin{subfigure}[t]{0.2\textwidth}
		\includegraphics[scale=0.75]{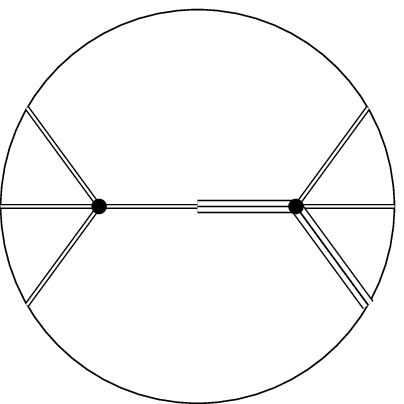}
		\caption{}
		\label{R4ex1b}
	\end{subfigure}	
~~~
	\begin{subfigure}[t]{0.2\textwidth}
		\includegraphics[scale=0.75]{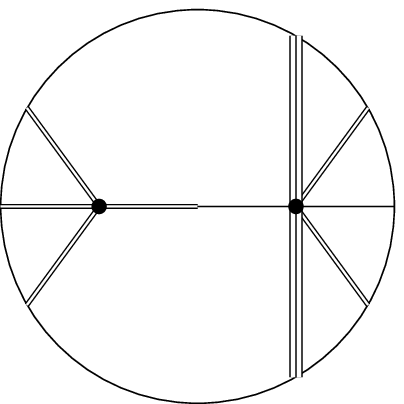}
		\caption{}
		\label{R48u}
	\end{subfigure}
~~~
	\begin{subfigure}[t]{0.2\textwidth}
	\includegraphics[scale=0.75]{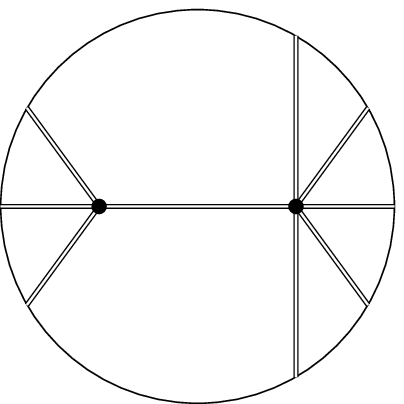}
		\caption{}
		\label{R48x}
	\end{subfigure}		
	\caption{Some connected functions of order $\epsilon^3$. In particular, two-vertex convolutions of 1PI  functions of order $\epsilon$ and $\epsilon^2$. }
	\label{2vd}
\end{figure}

Figure~\ref{2vd}~depicts some examples from this class.  Without loss of generality, we will focus on~\cref{R4ex1b}.  
 Let us begin here by recalling the tensorial structure of a  Riem$^2$ 
contraction  in~\cref{first-decomposition}. By direct analogy,  a Riem$^3$ contraction in the high-momentum regime can always be expressed as
\begin{align}\label{second-decomposition}
\tensor{R}{_{ab}^{cd}}\tensor{R}{_{mn}^{ab}}\tensor{R}{^{mn}_{cd}} &\;=\; \mathcal{X}^{pqcd}R_{abpq}\mathcal{X}^{abrs}R_{mnrs}\mathcal{X}^{mntu}R_{tucd}\;.
\end{align}

The expansion of~\cref{second-decomposition} is greatly simplified by the 
 fact that, in this regime, derivatives contract with derivatives and gravitons, with gravitons.  Having this in mind, one finds that a 1PI 4-point function takes on the generic form
\begin{align}\label{2decomposed41PI}
\avg{h^4}^{\epsilon^2}_{1PI} &\;=\; \frac{1}{2^3}\nabla_a\nabla^b h_{mn} \nabla_c\nabla^a h^{mq} \nabla_b h_{pq} \nabla^c h^{pn}\;.
\end{align}

The convolution of interest can then be written in coordinate space as
\begin{align}
\avg{h^6}^{\epsilon^3, (5a)}_{con} &\;=\; \avg{h^4}^{\epsilon}_{1PI} \cdot \avg{h^4}^{\epsilon^2}_{1PI}, \notag \\
		&\;=\;\int dr \int d\bar{r} \int d^4x \int d^4\bar{x} \; \; \nabla_i h(r,x) \nabla^i h(r,x) \nabla_j h(r,x) \avg{\nabla^j h(r,x) \bar{\nabla}_a \bar{\nabla}^b h(\bar{r},\bar{x})} \notag \\ \; \; \; \; & \;\times\; \bar{\nabla}_c \bar{\nabla}^a h(\bar{r},\bar{x}) \bar{\nabla}_b h(\bar{r},\bar{x}) \bar{\nabla}^c h(\bar{r},\bar{x})\;.
\end{align} 
As before, one has to integrate over both the $(r,x)$ and $(\bar{r},\bar{x})$ coordinates while considering the symmetrization of the external gravitons. The
 resulting expression  works out to
\begin{align}
\avg{h^6}^{\epsilon^3, (5a)}_{con} &\;=\; \frac{38}{9}\int dr \int d^4x \; \; \nabla_i h(r,x) \nabla^j h(r,x) \nabla^i\nabla_j h(r,x) \nabla_c h(r,x)\nabla^b h(r,x) \nabla_b \nabla^c h(r,x)\;,
\end{align}
which has the expected  eight-derivative signature of an $\epsilon^3$-order correlation function.

Also like before, when dealing with the tensor structure of the convolution, one can disregard the role of the derivatives. And because of this simplification, there is no distinction between once- and twice-differentiated gravitons, as far as the tensor structure is concerned. With this in mind, the tensor-structure part
of the  calculation entails the contraction of  of~\cref{1decomp1PI,2decomposed41PI} (but with their derivatives stripped), which yields
\begin{align}
\avg{h^6}^{\epsilon^3, (b)}_{con} &\; =\; \frac{1}{2^3} h^{mq}h^{pn}h_{pq} \avg{h_{mn}h_{eb}} h_{af} h_{di} h_{cj}\mathcal{X}^{abcd} \mathcal{X}^{efij}\;.
\end{align}  

To proceed, one  expands the $\mathcal{X}$'s, applies the graviton propagator in momentum space as per~\cref{GravPropagator},  discards terms with scalars and further
simplifies. After all this, the outcome is
\begin{align}
\avg{h^6} &\;=\; \frac{1}{2^4}h^{mq}h_{pq}h^{pn}h_{mn}h^{af}h_{af}\;.
\end{align}
A similar calculation can be performed for any of the other diagrams in~\cref{2vd}.
%
\begin{figure}[H]
\centering
\begin{subfigure}[t]{0.17\textwidth}
	\includegraphics[scale=0.75]{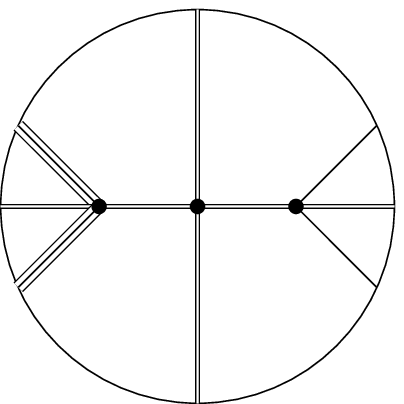}
	\caption{}
	\label{R48a}
\end{subfigure}
~~~
\begin{subfigure}[t]{0.17\textwidth}
	\includegraphics[scale=0.75]{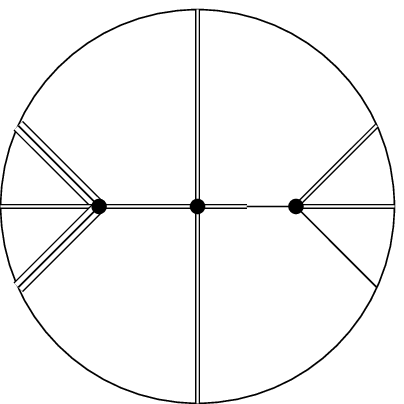}
	\caption{}
	\label{R48b}
\end{subfigure}
~~~
\begin{subfigure}[t]{0.17\textwidth}
	\includegraphics[scale=0.75]{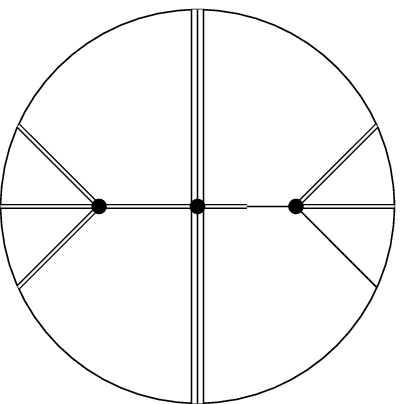}
	\caption{}
	\label{R48c}
\end{subfigure}
~~~
\begin{subfigure}[t]{0.17\textwidth}
	\includegraphics[scale=0.75]{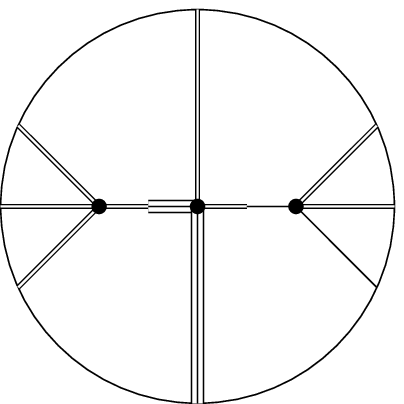}
	\caption{}
	\label{R48d}
\end{subfigure}
~~~
\begin{subfigure}[t]{0.17\textwidth}
	\includegraphics[scale=0.75]{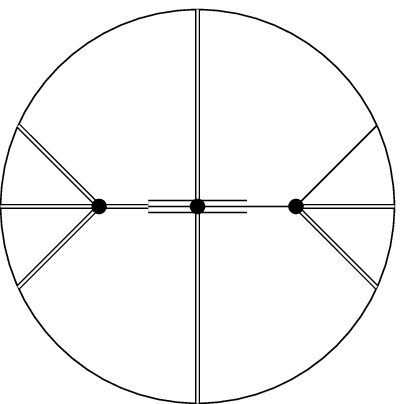}
	\caption{}
	\label{R48e}
\end{subfigure}
~~~
\begin{subfigure}[t]{0.17\textwidth}
	\includegraphics[scale=0.75]{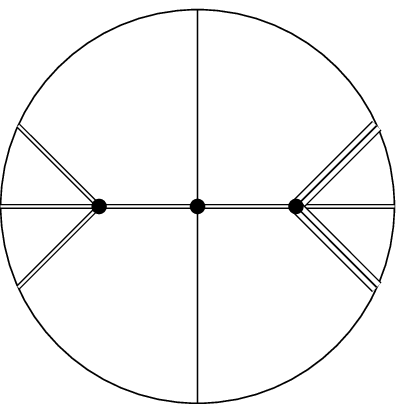}
	\caption{}
	\label{R48f}
\end{subfigure}
\caption{Some three-vertex diagrams at order $\epsilon^3$.}
\label{All3vR4}
\end{figure}

\subsection{Higher order connected diagrams}
\begin{figure}[H]
\centering
\includegraphics[scale=0.75]{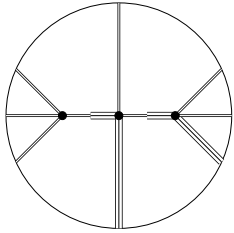}
\caption{An example of a connected $8$-point function of order $\epsilon^6$.}
\label{R7example}
\end{figure}

As  should by now be evident,  there is an increasingly diverse set of connected diagrams with increasing orders of $\epsilon$.  In addition to the different ways of distributing the $\epsilon$'s at any given order, one can always  obtain higher-point amplitudes either  by adding undifferentiated gravitons from  both the metric determinant and the contravariant metrics or by inserting  Einstein diagrams ({\em i.e.}, extra vertices) at any point along the chain. In short, as the perturbative order increases, the combinatoric calculations grow exponentially fast in their complexity. For instance,~\cref{All3vR4} shows just a sample of the many distinct $8$-point functions that emerge when \textit{only} three 1PI functions of  order $\epsilon^0$, $\epsilon^1$ and $\epsilon^2$ are convolved. One can construct even more three-vertex, connected $8$-point functions up to an order of  $\epsilon^6$. (Higher orders than this would violate the high-momentum rulebook.) 
As an example, three 1PI functions of orders $\epsilon$, $\epsilon^2$ and $\epsilon^3$ can be used to construct the diagram in~\cref{R7example}.

And so, in spite of  there being only three types of lines --- gravitons carrying zero, one, or  two  derivatives --- there are still  many ways of combining them 
to obtain  a large number of vertices. At present, we do not have a general formula for determining the coefficients for connected functions with an arbitrary number of vertices at an arbitrary order. However, it is always possible, at least in principle,  to make a case-by-case study of connected diagrams at any specified order.

\section{Experimental considerations}

For the sake of making contact with experiment, it is useful to reexpress the high-momentum regime from the perspective of the boundary gauge theory. 
 (Also see~\cite{Brustein:2012dg} for additional elaboration.) First recall that, in the bulk, this regime is defined by
\begin{align}\label{HMcond2}
1 \;\ll\; \omega L \;\ll \; TL\;.
\end{align}  

The AdS boundary naturally inherits  a planar geometry from that of the black brane. And yet, in spite of this, the boundary can be viewed as having an $S^3$ geometry~\footnote{The validity of this viewpoint depends on the boundary theory being conformal, at least up to some approximation.} with a  radius $\cal{R}$  that is determined by  the AdS curvature scale, $\mathcal{R} \sim L$~\cite{Witten:1998zw}. The radius $\mathcal{R}$ should, on the other hand, scale with the 
spatial extent of the gauge-theory fluid~\cite{Friess:2006kw}.  Then, since $T$ is the temperature of both the black brane and the fluid, the regime can be redefined as
\begin{align}\label{holographicHMregime}
1\; \ll\; \omega \mathcal{R}\; \ll\; T\mathcal{R}\;.
\end{align}

Given access to experimental data on gauge-theory connected functions that is consistent with~\cref{holographicHMregime}, one could use our framework as a means for identifying the gravitational dual to the QGP (at least in principle). To understand how this can work, let us consider the following schematic forms for the correlation functions of the gauge theory:~\footnote{The highest-order term in any given line is the highest  order that would not be violating  the rules of  high-momentum regime.}
\begin{align}\label{stress-to-graviton-multi}
& \avg{T_{xy}T_{xy}}\; =\; \mathcal{A}_0 \epsilon^0 \avg{h_{xy}h_{xy}} + \mathcal{A}_1 \epsilon^1 \avg{h_{xy}h_{xy}}\;, \notag \\
& \avg{T_{xy}^4}\; =\; \mathcal{A}_0 \epsilon^0 \avg{h_{xy}^4} + \mathcal{A}_1\epsilon \avg{h_{xy}^4} + \mathcal{A}_2\epsilon^2 \avg{h_{xy}^4} + \mathcal{A}_3\epsilon^3 \avg{h_{xy}^4}\;, \notag \\
& \avg{T_{xy}^6}\; =\; \sum_{i=0}^{5} \mathcal{A}_i\epsilon^i \avg{h_{xy}^6}\;, \notag \\
& \; \; \vdots \; \; \; \vdots \; \; \; \; \; \; \;\; \;\; \; \; \; \; \vdots \; \; \vdots  \notag \\
& \avg{T^{2n}_{xy}}\; =\; \sum_{i=0}^{2n-1} \mathcal{A}_i \epsilon^i \avg{h_{xy}^{2n}}\;.
\end{align} 
Here, the left-hand side in all cases is meant to be an experimentally 
measured  number and the $\mathcal{A}_i $'s are unknown constants that must be 
 determined if one is to identify the gravitational dual. We are, however, 
free to fix  $\mathcal{A}_0 = 1$ and regard the other $\mathcal{A}$'s as ratios
of this one.  Ideally, one would want to have $k$ linear equations with $k$ unknowns. However, as can be seen in~\cref{stress-to-graviton-multi}, every $(2n+2)$-point function has two more unknowns on the right-hand side than did the previous $2n$-point function. But one could still circumvent this impediment if the experimental apparatus was not  sensitive to all of the $2n$-point functions on the right-hand side. That this is indeed plausible will be shown next. 

Let us first recall that every Riemann tensor draws one of $\omega^2$, $\omega k$ or $k^2$, which will be collectively labelled as $s$ (with reference to the center-of-mass energy $\sqrt{s}$). This allows us to rewrite the gravitational Lagrangian schematically as ({\em cf}, Eq.~(\ref{Lagrangian0}))
\begin{align}
\mathcal{L} &\;=\; s \sum_{i=0} \frac{a_i\ell^i}{\mathcal{R}^2}\;,
\end{align}  
where $\ell = \epsilon s \mathcal{R}^2$ and the $a$'s are unknown coefficients. Notice that $\ell \sim \frac{s}{M_P^2}$, where $M_P$ is the Planck mass, and so it can be expected that $\ell < 1$. 

The high-momentum regime can now be recast as
\begin{align}\label{hm10}
\ell_{\text{min}}\; \ll\; \ell \; \ll \; \ell_{\text{max}}\;,
\end{align}
with $\ell_{\text{min}} \sim \epsilon$ and $\ell_{\text{max}} \sim \epsilon(T\mathcal{R})^2$. Depending on the (dimensionless)  sensitivity $W$ of the experimental apparatus, one could hope to find an $\ell$ in the high-momentum regime such that $\ell^k < W$ but $\ell^{k-1} > W$, for some suitably small integer $k$. This would then allow for the elimination of terms at order $\ell^k$ and greater, thus reducing the unknown coefficients to the range  $0 \leq i < k$. One could then solve for the $a_i$'s since $\mathcal{A}_i = \mathcal{A}_i(a_0=1,a_1,\dots,a_{i-1},a_i)$.

\section{Conclusion}

To summarize,   we have been considering  a generic higher-derivative theory of gravity living in an
asymptotically AdS,  black brane  spacetime.  The main purpose was to show how to translate a certain
class of 1PI graviton scattering amplitudes into connected  amplitudes at the AdS boundary.  And so, by
way of the  gauge--gravity and  ``fluid--black brane''  dualities, what we  have  effectively calculated
are   stress-tensor correlation functions for  a strongly coupled fluid  whose pedigree is  the
(presumptive)  gauge-theory  dual of the AdS brane theory.  The   amplitude calculations were carried out in a particular kinetic
regime of large momenta and large scattering angles.~\footnote{The large scattering angles can be seen
from a calculation of the Mandelstaam variables for the case of a four-graviton scattering problem~\cite{Brustein:2012he}.} This regime is particularly  well suited  for distinguishing terms in a higher-derivative theory of
gravity, as it assigns added weight to the  higher-derivative contributions (relative to those of
Einstein's theory), which would otherwise be perturbatively suppressed as per general arguments from
string theory.  This fits nicely with  our ultimate objective, which is to  provide a means for using
experimental data to identify the gravitational dual to the QGP.  However, for such a scheme to work, it
is necessary that the regime is experimentally accessible and that the window of accessibility provides
large-enough momenta to achieve the desired task. Nevertheless, there are some  reasons for optimism
\cite{Brustein:2012dg}.

In most studies on Au+Au collisions, such as those at RHIC, a lot of information pertaining to certain species of particles is usually discarded when it comes to the calculation of correlation functions. In many cases, it is because only lower-point functions, such as the $2$- and $4$-point functions, are necessary for some particular calculation. However, a recent study in~\cite{Lin:2017xkd} shows that the measurement of higher-order multi-point functions provides a much clearer picture for  all the relevant  species of  particles. Meanwhile, as we have argued here, the inclusion of such higher-point functions could some day lead to a finely tuned picture of the dual to the QGP. On both this basis and that of~\cite{Lin:2017xkd}, it is our contention that the higher-point functions should not be so
recklessly disregarded.

\section*{Acknowledgements}
The research of AJMM received support from an NRF Incentive Funding Grant 85353 and NRF Competitive Programme Grant 93595. MMWS is supported by an NRF bursary through Competitive Programme Grant 93595 and a Henderson Scholarship from Rhodes University.

\bibliographystyle{unsrtnat}

\end{document}